\documentclass{iaus}
\usepackage{graphicx}

\title[Magnetic Fields in Large-Scale Structure] 
{Detecting Magnetic Fields in Large-Scale Structure with Radio Polarization}

\author[Brown, Rudnick \& Farnsworth]   
{Shea Brown$^1$, Lawrence Rudnick$^1$ \break \and Damon Farnsworth$^1$}

\affiliation{$^1$Department of Astronomy, University of Minnesota,
Minneapolis, MN 55455, USA}

\pubyear{2009}
\volume{259}  
\pagerange{119--126}
\date{?? and in revised form ??}
\jname{Cosmic Magnetic Fields: from Planets, to Stars and Galaxies}
\editors{K.G. Strassmeier, A.G. Kosovichev \& J. Beckmann, eds.}
\begin{document}

\maketitle

\begin{abstract} We present our attempts to detect magnetic fields in 
filamentary large-scale structure (LSS) by observing polarized synchrotron 
emission emitted by structure formation shocks. Little is known about the 
strength and order of magnetic fields beyond the largest clusters of 
galaxies, and synchrotron emission holds enormous promise as a means of 
probing magnetic fields in these low density regions. We report on 
observations taken at the Green Bank Telescope which reveal a possible Mpc 
extension to the Coma cluster relic. We also highlight the major obstacle 
that diffuse galactic foreground emission poses for any search for 
large-scale, low surface-brightness extragalactic emission. Finally we 
explore cross-correlation of diffuse radio emission with optical tracers 
of LSS as a means to statistically detecting magnetic fields in the 
presence of this confounding foreground emission.

\end{abstract}

\firstsection 

\vspace{-5mm} \section{Introduction} One of the goals of the next 
generation(s) of radio telescopes (LOFAR, MWA, EVLA, SKA) will be to 
determine the origin of cosmic magnetism. As noted by \cite{donn08}, the 
large amplification of fields to several $\mu$G in rich galaxy cluster 
cores largely erases the signatures of their origins, whereas the 
0.1$\mu$G fields expected in filaments can retain indicators of their 
origins.  Thus, it is critical to study the strength and structure of 
filament fields. Radio synchrotron emission is a powerful means of 
detecting magnetic fields in these regions, and shocks from infall into 
and along the filamentary structures between clusters are now widely 
expected to generate relativistic plasmas (e.g., \cite[Ryu et al. 
2008]{Ryu08}, \cite[Skillman et al. 2008]{skil08}). We are searching for 
the synchrotron signatures of these shocks.

\vspace{-6mm} \section{Results}\label{sec:results} The Coma cluster's 
radio relic is evidence of shock activity due to infall from a smaller 
cluster. We performed 1.4 GHz observations of this cluster with the Green 
Bank Telescope (GBT). Fig. 1 shows a Stokes I image of the Coma region 
after subtracting point sources and applying a 20$^{\prime}$' median 
weight filter. Diffuse emission, both polarized (not shown) and 
unpolarized, extends a full 2~Mpc aligned with the relic. It is unclear if 
this is all related to the shock/filament, though the fact that it is 
polarized is consistent with shock compression. We are currently analyzing 
350~MHz RM-Synthesis (\cite[Brentjens \& de Bruyn 2005]{bren05}) data from 
the WSRT which could resolve the ambiguity.

We have also identified high galactic-latitude ``Faraday screens" as a 
significant source of confusion for low frequency polarization studies 
(Brown, Rudnick, \& Farnsworth in prep). These Faraday rotating clouds 
modulate the smooth galactic background to produce polarized power in an 
interferometer that has no total-intensity counterpart (Fig. 1). These 
features will likely be ubiquitous at frequencies $<$ 200~MHz, and will 
present a foreground contamination to extragalactic studies. Intrinsic 
angle changes in galactic emission, independent of Faraday rotation, will 
also add polarized power into an interferometer while the total intensity 
remains invisible because it is smooth on large scales.

 Motivated by these foreground issues, we are exploring cross-correlation 
methods for detecting synchrotron emission in filamentary LSS. Similar to 
detections of the Integrated Sachs-Wolfe effect (ISW), we cross-correlate 
large FOV radio maps with optical/IR tracers of LSS. As a first step, we 
performed a zero-shift cross correlation of a 34$\times$34 degree area of 
the 1.4~GHz Bonn survey with the corresponding distribution of 2MASS 
galaxies (0.03 $<$ z $<$ 0.04). To assess the significance of a positive 
correlation, we also correlated the 2MASS galaxies with 24 random fields 
of the same size from the rest of the Bonn survey. Though we obtained a 
null result (as expected), we found that adding a signal weighted by the 
2MASS image with a mean surface brightness of $\sim$1~mK was sufficient to 
produce a 3$\sigma$ƒ positive correlation. This injected signal is below 
the rms noise of the Bonn survey and demonstrates the power of this 
technique.

\begin{figure}
 \includegraphics[width=6.1cm]{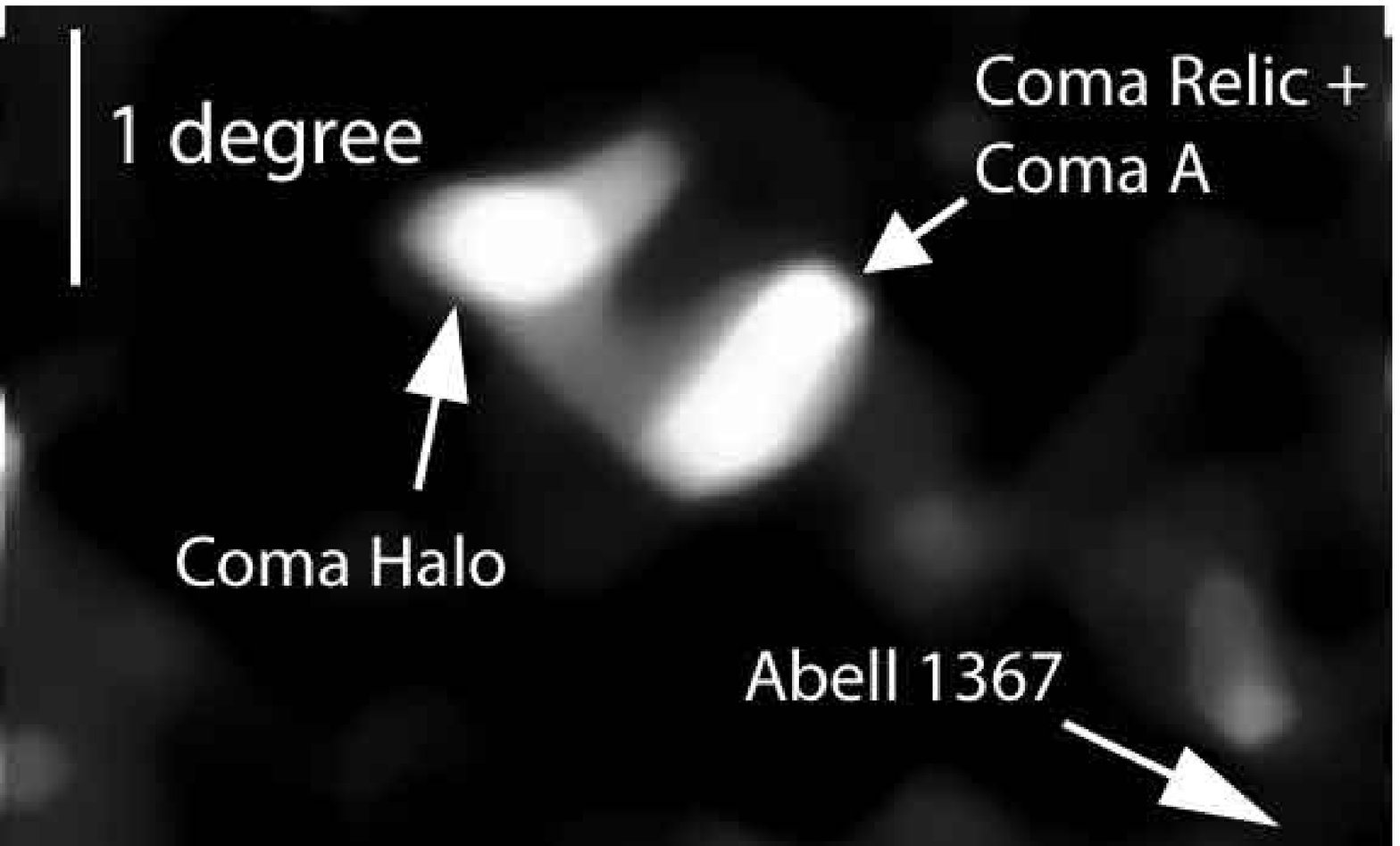}
 \includegraphics[width=7.9cm]{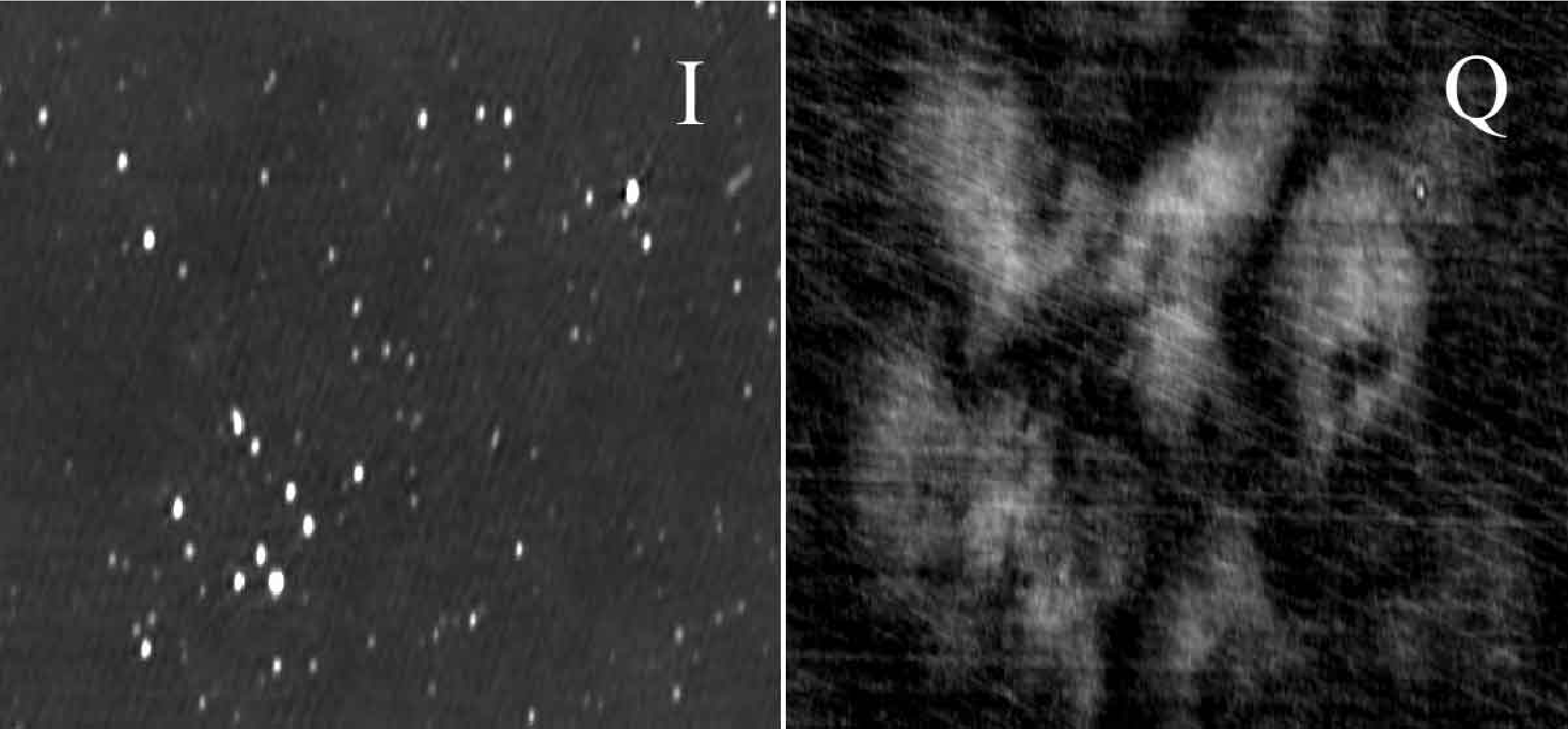}
  \caption{
Left:  GBT Stokes I image of Coma cluster; Center/right: Westerbork 
Synthesis Radio Telescope (WSRT) Stokes I/Q images of field north of Coma 
cluster where \cite{kron07} report diffuse source ``B".  We do not confirm 
this source in I, although Faraday modulated Q and U patches are seen in 
this region.
}\label{fig:grey} 
\end{figure}

\vspace{-6mm}
\section{Conclusions}\label{sec:concl} Measuring the weak magnetic fields 
within filaments of galaxies is critical to determining the origins of 
cosmic magnetism, and synchrotron radio emission can be a powerful tracer 
of magnetic fields in these regions. Diffuse Galactic synchrotron emission 
and Faraday rotating plasma will present a significant foreground problem 
for upcoming low-frequency surveys, particularly in polarization. We 
demonstrate that cross-correlation of synchrotron maps and tracers of 
large-scale structure provides a powerful method for detecting faint 
emission in filaments.

\vspace{-3mm} \begin{acknowledgments} The GBT is a facility of the 
National Science Foundation, operated by NRAO under contract with AUI, 
Inc. The WSRT is operated by the ASTRON (Netherlands Foundation for 
Research in Astronomy) with support from the Netherlands Foundation for 
Scientific Research (NWO). Partial support for this work at the University 
of Minnesota comes from the U.S. National Science Foundation grant 
AST~0607674.

\end{acknowledgments}


\vspace{-5mm}
\begin{thebibliography}{}

\bibitem[Brentjens \& de Bruyn (2005)]{bren05} Brentjens, M.~A., \& de
Bruyn, A.~G. 2005, \textit{A\&A}, 441, 1217

\bibitem[Donnert et al. (2008)]{donn08} Donnert, J., Dolag, 
K., Lesch, H., M$\ddot{u}$ller, E. 2008, arXiv:0808.0919

\bibitem[Kronberg et al. (2007)]{kron07} Kronberg, P.~P.,
Kothes, R., Salter, C.~J., \& Perillat, P. 2007, \textit{ApJ}, 659, 267

\bibitem[Ryu et al. (2008)]{Ryu08} Ryu, D., Kang, H., Cho, 
J., \& Das, S. 2008, Science, 320, 909

\bibitem[Skillman et al. (2008)]{skil08} Skillman, S.~W., 
O'Shea, B.~W., Hallman, E.~J., Burns, J.~O., 
\& Norman, M.~L. 2008, \textit{ApJ}, 689, 1063

\end{thebibliography}
\end{document}